\documentclass[10pt, conference, compsocconf]{IEEEtran} 

\usepackage[pdftex]{graphicx}
\usepackage{bibspacing}
\usepackage{cite}
\usepackage{xcolor}
\usepackage{mdwmath}
\usepackage{mdwtab}
\usepackage[caption=false,font=footnotesize]{subfig}

\makeatletter
\newcommand*\titleheader[1]{\gdef\@titleheader{#1}}
\AtBeginDocument{%
	\let\st@red@title\@title
	\def\@title{%
		\bgroup\normalfont\large\centering\@titleheader\par\egroup
		\vskip1.5em\st@red@title}
}
\makeatother

\title{QoE-Aware Cross-Layer Architecture for Video Traffic over Internet}
\titleheader{\textcolor{blue}{Published in: 2014 IEEE REGION 10 SYMPOSIUM. Date Added to IEEE Xplore: 24 July 2014. Date of Conference: 14-16 April 2014\\
		\textcopyright 2014 IEEE.  Personal use of this material is permitted.  Permission from IEEE must be obtained for all other uses, in any current or future media, including reprinting/republishing this material for advertising or promotional purposes, creating new collective works, for resale or redistribution to servers or lists, or reuse of any copyrighted component of this work in other works.\\
		DOI: 10.1109/TENCONSpring.2014.6863089}}

\author{\IEEEauthorblockN{Safeen Qadir$^1$, Alexander A. Kist$^1$ and Zhongwei Zhang$^2$}
	\IEEEauthorblockA{$^1$ School of Mechanical and Electrical Engineering\\
		\{safeen.qadir, kist\}@ieee.org\\
		$^2$ School of Agricultural, Computational and Environmental Sciences\\
		zhongwei.zhang@usq.edu.au\\
		University of Southern Queensland, Australia\\
}}

\begin{document}

\maketitle

\thispagestyle{plain}
\pagestyle{plain}

%
\title{QoE-Aware Cross-Layer Architecture for Video Traffic over Internet}



\maketitle

\begin{abstract}
The emergence of video applications and video capable devices have contributed substantially to the increase of video traffic on Internet. New mechanisms recommending video rate adaptation towards delivering enhanced Quality of Experience (QoE) at the same time making room for more sessions. This paper introduces a cross-layer QoE-aware architecture for video traffic over the Internet. It proposes that video sources at the application layer adapt their rate to the network environment by controlling their transmitted bit rate dynamically; and the edge of network at the network layer protects the quality of the active video sessions by controlling the acceptance of new session through a video-aware admission control. In particular, it will seek the most efficient way of accepting new video session and adapting transmission rates to free up resources for more session while maintaining the QoE of active sessions. The proposed framework will contribute to the preparation for the extreme growth of video traffic in the foreseeable future. Simulation results show that the proposed cross-layer architecture guarantees the QoE for the admitted sessions and utilizes the link more efficiently comparing to the rate adaptation only architecture.
\end{abstract}


\IEEEpeerreviewmaketitle

\section{Introduction}
The emergence of video applications and video capable devices have contributed substantially to the increase of video traffic on the Internet. Cisco forecasts that "The number of devices connected to IP networks will be nearly three times as high as the global population in 2016. There will be nearly three networked devices per capita in 2016, up from over one networked device per capita in 2011" \cite{Cisco2012}. Non-PC devices will generate 31 percent of IP traffic by 2016 growing from 22 percent in 2011 \cite{Cisco2012}. The high quality video playback capability of most of these devices is a key driver of the evolution of new mechanisms recommending video rate adaptation towards delivering enhanced Quality of Experience (QoE) at the same time making room for more sessions.

The massive demand for video anytime and anywhere has led to the development of adaptive streaming solutions that are able to deliver video with a maintained QoE. QoE is a measure of the user perceived quality of a network service. One of these mechanisms which delivers video over Internet through web browser is the HTTP Adaptive Streaming (HAS) \cite{Oyman2012} by which a client and the web/media server decide on which rate they should communicate. Many companies have introduced their HAS solutions such as Microsoft Smooth Streaming, Apple HTTP Live Streaming and Adobe HTTP Dynamic Streaming.

Admission control is a well known technique to keep the traffic load at an acceptable level and guarantee the Quality of Service (QoS) for the admitted flows. This idea has been adopted in the past in QoS architectures such as Diffserv. The IETF recently standardized a simple, robust and scalable measurement-based admission control and flow termination for Diffserv domains based on Pre-Congestion Notification (PCN) \cite{EardleyP.2009}.

Enhancement can be done towards improving the QoE on all layers (from video encoding to decoding) and across the access and/or core networks. Cross-layer optimization has been proposed recently to provide QoE by the cooperation of several layers in the protocol stack. It indicates a combined enhancement in more than one layer of Open Systems Interconnection (OSI) model to optimize the audio/video quality, throughput or QoE. It enables communication and interaction between layers by allowing one layer to access the data of another layer. For example, having knowledge of the current available bandwidth (the network layer) will help a source to perform the rate adaptation (the application layer) to optimize the throughput or QoE.

Scalable video encoding techniques have been proposed to cope with the problem of Internet resource uncertainty and support different devices. The Scalable Video Coding (SVC) extension of the H.264/AVC standard from the joint video team of the ITU-T Video Coding Experts Group (VCEG) and the ISO/IEC Moving Picture Experts Group (MPEG) provides the transmission and decoding support of video partial bit streams to different applications and devices. It enables lower temporal or spatial resolutions or reduced quality while retaining a reconstruction quality that is high relative to the rate of the partial bit streams \cite{Schwarz2007}. The problem with the SVC however, is that the bit rates can not be changed on the fly. 


Although dynamic rate adaptation enhances video quality, accepting more sessions than a link can accommodate will degrade the quality. We studied how implementing the rate-adaptation by a video source can maintain a better QoE over non-adaptive source in \cite{Qadir2013}. However, the friendly behavior of Internet's transport protocol (TCP) accommodates every video session and makes room for everyone. This causes degradation of the QoE of all video sessions in a bottleneck link. That is because for a large number of video sessions, the adaptive source attempts to adapt the transmission rate of all video sources in order to share the available link capacity without considering how much the received QoE will be affected by the adaptation process. Therefore, along with the adaptable video source there is a need for a mechanism to control the number of video sessions which can be accommodated with acceptable QoE and to protect the QoE of the current sessions.

This paper proposes a cross-layer QoE optimization for video traffic over Internet. It addresses the problem of QoE degradation in a bottleneck network. In particular, it allows video sources at the application layer to adapt themselves to the network environment by controlling their transmitted bit rate dynamically, and the edge of network at the network layer to protect the quality of active video sessions by controlling the acceptance of new session through a proposed video-aware admission control. 

The application layer contributes to the optimization process by dynamically adapting source bit rate based on the condition of network and the network layer controls admission of new video session based on the aggregated video traffic.

\section{Related Work}
There are a number of studies that considering cross-layer optimization for the sake of video quality enhancement, such as \cite{Duong2010}, \cite{Guerses2005} and \cite{Gross2004}, or throughput improvement such as \cite{Shabdanov2012}. We will include only those which aimed at QoE. \cite{Khalek2012} introduced an Application/MAC/Physical (APP/MAC/PHY) cross-layer architecture that enables optimizing perceptual quality for delay-constrained scalable video transmission. Using the acknowledgment (ACK) history and perceptual metrics, an online mapping of QoS-to-QoE has been proposed to quantify the packet loss visibility from each video layer. A link adaptation technique that uses QoS-to-QoE mapping is developed at the PHY layer to provide perceptually-optimized unequal error protection for each video layer according to packet loss visibility. While at the APP layer, a buffer-aware source adaptation is proposed by which the sender's rate is adapted by selecting the set of temporal and quality layers without incurring playback buffer starvation based on the aggregate channel statistics. To avoid re-buffering and frame freeze, a video layer-dependent retransmission technique per packet at the MAC layer limits the maximum number of packet retransmission based on the packet layer identifier. The next retransmission of packet is given a lower order of Modulation and Coding Scheme (MCS). The article concluded that the proposed architecture prevents playback buffer starvation, handles short-term channel fluctuations, regulates the buffer size, and a 30\% increase in video capacity is achieved compared to throughput-optimal link adaptation. Apart from the encoding implementation using JSVM, experimental or simulation is not carried out to study the joint coherence of the proposed techniques. In addition to its limitation to the SVC, the study didn't target a specific underlying wireless technology.

A cross-layer scheme for optimizing resource allocation and user perceived quality of video applications based on the QoE prediction model that maps between object parameters and subject perceived quality proposed by \cite{Ju2012}. \cite{Fiedler2009} promoted automatic feedback of end-to-end QoE to the service level management for better service quality and resource utilization by presenting a QoE-based cross-layer design of mobile video systems. It discussed challenges of incorporating QoE concepts among different layers and suggested approaches span across layers such as efficient video processing, advanced realtime scheduling.

\cite{Politis2012} incorporated a rate adaptation scheme and the IEEE 802.21 Media Independent Handover (MIH) framework to propose a QoE-driven seamless handoff scheme. The rate is controlled by adapting the Quantization Parameter (QP) for single layer coding (H.264/AVC) and dropping the enhancement layers for scalable coding (H.264/SVC). The paper concluded that the proposed QoE-driven handover implemented in a real test-bed outerperforms the typical Signal-to-Noise Ratio (SNR) based handover and improves the perceived video quality significantly for both coding. However it can be better maintained with the H.264/SVC. The study is merely a comparison between the two coding techniques for maintaining the QoE of wireless nodes during the handover process.

\cite{Khan2006} proposed an application-driven objective function that jointly optimizes the application layer, data-link layer and physical layer of the wireless protocol stack for video streaming. The proposed cross-layer optimizer periodically receives information from the video server and selects the optimal parameter settings of the different layers based on the outcome of the maximization of an object function that depends on the reconstruction quality in the application layer. The parameters that can be optimized are the source rates at the application layer and modulation schemes, Binary Phase Shift Keying (BPSK) (total rate of 300kb/s) or Quaternary PSK (QPSK) (a total rate of 600 kb/s) at the radio link layer (radio link layer = physical + data link layer). 

\cite{Thakolsri2009} extends \cite{Khan2006}'s work from application-driven to a QoE-based cross-layer design framework for High Speed Downlink Packet Access (HSDPA) to maximize user satisfaction. It combines both capabilities of the HSDPA link adaptation and multimedia applications rate adaptation. Relevant parameters from the radio link and application layers are communicated to a cross-layer optimizer. The optimizer acts as a downlink resource allocator which periodically reviews the total system resources and makes an estimate of the time-share needed for each possible application-layer rate. It re-adapts the application rate if necessary. The proposed QoE-based cross layer optimized scheme was simulated using OPNET and compared to both the throughput optimized and non-optimized HSDPA systems. The paper found that the user perceived quality was significantly improved compared to other two systems. The study made use of the adaptability feature of HAS and aggressive TCP to control the application rates. Another shortcoming is that the Mean Opinion Score (MOS) was defined as a function of the transmission rate only.

\cite{Latre2011} proposed several techniques to optimize the QoE in multimedia network in terms of the number of admitted sessions and video quality. Traffic adaptation, admission control and rate adaptation were proposed within an automatic management layer using both simulation and emulation on a large-scale testbed. The study focused on multimedia services such as Internet Protocol TV (IPTV) and network-based Personal Video Recording (PVR).
The viability of implementation was investigated using neural network and was compared with an analytical model. The study shown that the proposed QoE optimizing techniques can successfully optimize the QoE of multimedia services.

The research discussed above proposed rate adaptation for layered video such as the SVC. The video content (base and enhancements layers) generated by the SVC are injected to network by video source, then network decides whether they are forwarded or dropped. In contrast, this paper proposes rate adaptation for single layer video. Instead of sending the whole video content to network, the video source based on the condition of the network decides at what rate has to transmit. By using this strategy, non-necessary data is not sent to the network and the network is not overcongested during congestion time.


Other than \cite{Latre2011}, none of the literatures has employed a combined rate adaptation and admission control in a cross-layer design for the QoE optimization. However our proposal is different from the extended-PCN admission control presented in \cite{Latre2011} as the video rate adaptation algorithm re-scales the rate of layered video flows.

\section{QoE-Aware Cross-Layer Architecture}
\begin{figure}[!t]
\centering
\includegraphics[width=\linewidth]{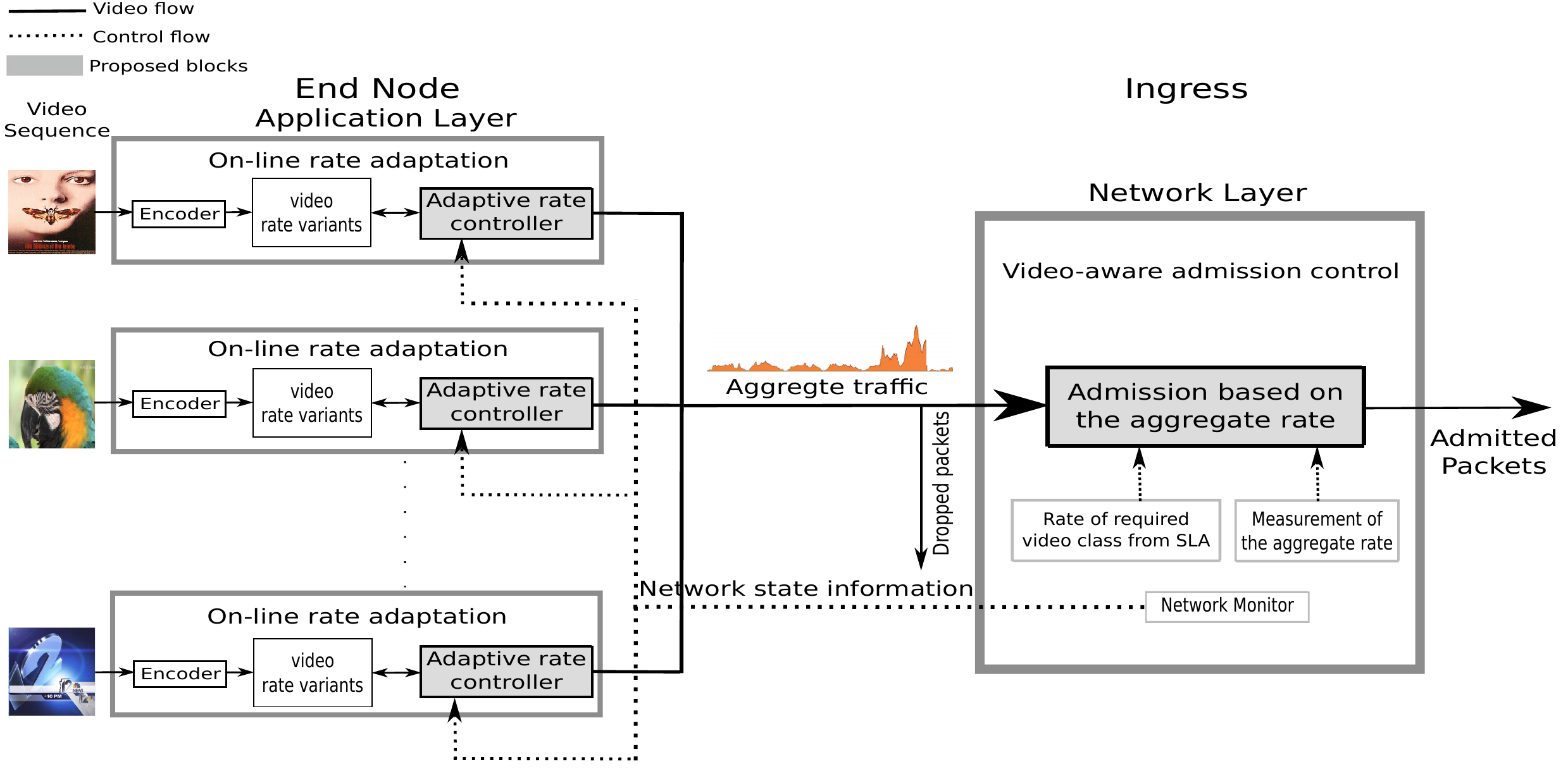}
\caption{The QoE-aware cross-layer architecture for video traffic over Internet}
\label{Proposed_Framework_paper3_conf}
\end{figure}

Figure \ref{Proposed_Framework_paper3_conf} shows the cross-layer architecture in which the proposed blocks are highlighted. The rate adaptation is performed at the application layer and admission control at the network layer. The proposed framework employs parameters from relevant layers; application and network layers in this paper. The key parameter to be considered for the cross-layer optimization is the source data rate from the application layer, while for the proposed admission control, the video flow identified by the source and destination IP address will be taken into account.

Encoders that provides quality variability such as MPEG-4 can be used to produce different video quality from the video scenes. The rate controller adapts the transmitting rate based on the load on the link. The load is monitored and estimated by the network monitor and the information is sent back to the rate controller via the acknowledgment packet of the TCP. The rate controller based on the information received from the network monitor on the network state selects a suitable video quality of available bit rates (video rate variants in Figure \ref{Proposed_Framework_paper3_conf}) for each Group of Picture (GoP). An open loop Variable Bit Rate (VBR) controller requires access to both video content and network state information. The Explicit Congestion Notification (ECN) bit in the acknowledgment packet of the TCP header can be utilized for the purpose of network monitoring. The rate controller at the sender side reduces its transmission rate by selecting a lower video rate variant if ECN 1 is detected in the acknowledgment packet.

Evalvid-RA \cite{Lie2008} is used to implement an on-line rate adaptation from different encoded videos each with valid range of QP from 1-31. High QP causes a high bit rate and better quality. The admission control mechanism measures the network load and based on that makes the admission decision. The rate that provides the required QoE and to which the user has subscribed through the Service Level Agreement (SLA) is added to the measured rate. The new requested session will be admitted only if the sum of the aggregate traffic rate on the link plus the video class rate is less than or equal to the link capacity.

\section{Simulation, Results and Discussion}
NS-2 \cite{ns2} is used to implement the proposed scheme of both architectures; rate-adaptation only (\emph{RA Only}) architecture and cross-layer (\emph{Cross-Layer}) architecture. In the \emph{RA Only} architecture, the video sources implement rate adaption only, while in the \emph{Cross-Layer} architecture, the ingress node implements an admission control mechanism in addition to the rate adaptation. 

The simulation was configured so that new session was requested randomly within every second of the simulation time. Maximum of 15 video sessions were competing for the bandwidth of a dumbbell topology in the simulations which run for the duration of 50 seconds. The 10 seconds "Akiyo" video sequence was encoded with different quantizer values using \cite{ffmpeg2004} to generate 30 different encoded bit rates. The Peak Signal-to-Noise Ratio (PSNR) was calculated from the simulation trace files and mapped to the MOS as the quality metric of the decoded videos. To investigate the performance of both architectures in terms of the QoE and number of the admitted sessions, the bottleneck link was configured with different bandwidths 2, 4, 6 and 9 Mbps. 

\begin{figure}[!t]
\centering
\includegraphics[width=\linewidth]{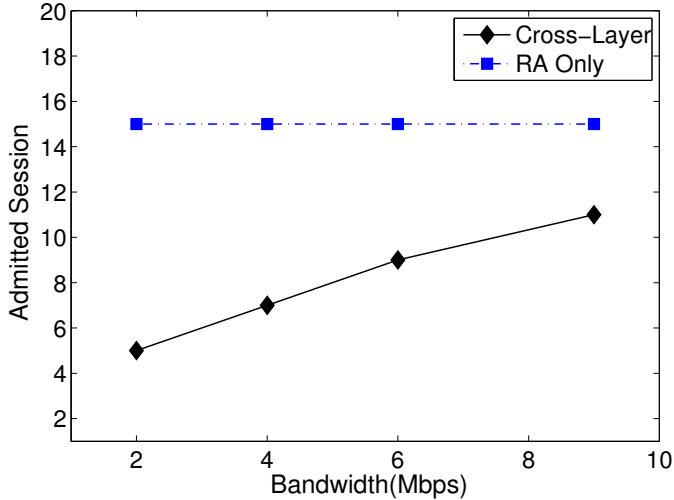}
\caption{Admitted sessions of the \emph{Cross-Layer} and \emph{RA Only} architectures for different bandwidth}
\label{session_bandwidth}
\end{figure}

\begin{figure}[!t]
\centering
\includegraphics[width=\linewidth]{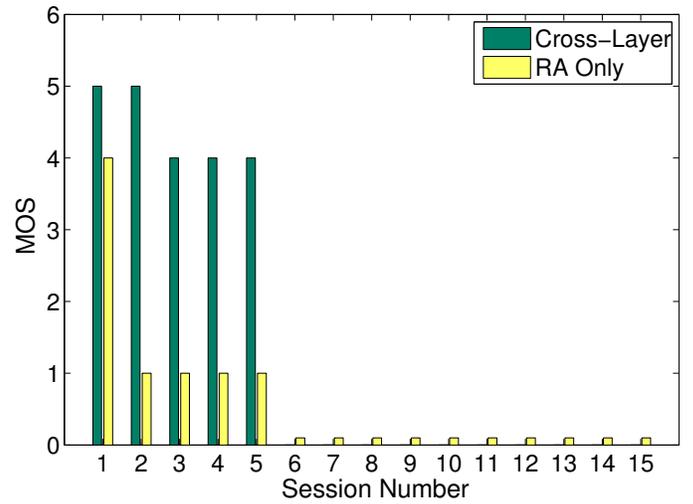}
\caption{The QoE of the admitted sessions for 2Mbps bandwidth}
\label{session_mos_2M}
\end{figure}

\begin{figure}[!t]
\centering
\includegraphics[width=\linewidth]{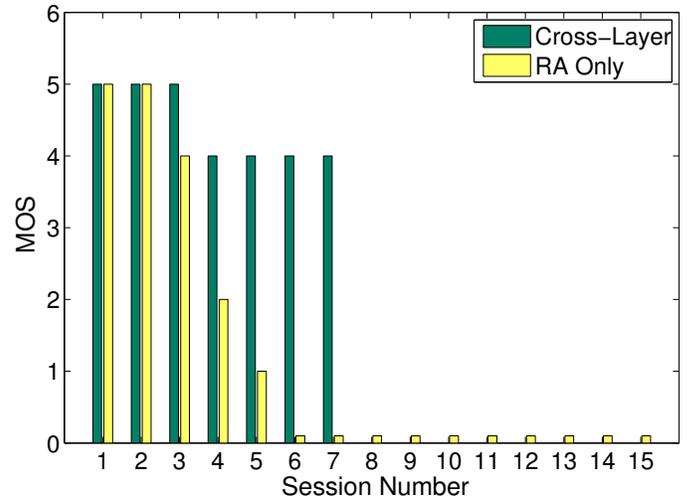}
\caption{The QoE of the admitted sessions for 4Mbps bandwidth}
\label{session_mos_4M}
\end{figure}

\begin{figure}[!t]
\centering
\includegraphics[width=\linewidth]{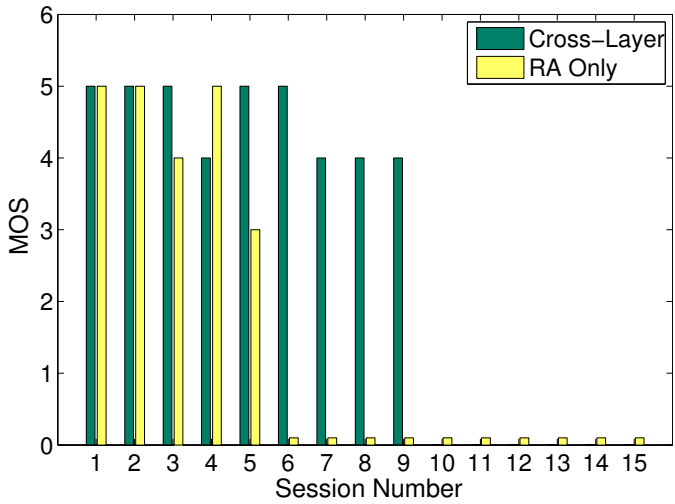}
\caption{The QoE of the admitted sessions for 6Mbps bandwidth}
\label{session_mos_6M}
\end{figure}

\begin{figure}[!t]
\centering
\includegraphics[width=\linewidth]{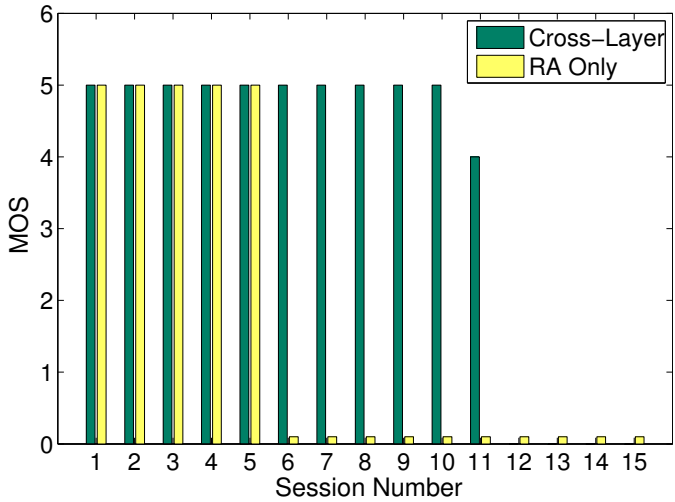}
\caption{The QoE of the admitted sessions for 9Mbps bandwidth}
\label{session_mos_9M}
\end{figure}

In the \emph{RA Only} architecture, any number of video session had been admitted, however for the sake of simplicity, the maximum number of admitted sessions was limited to 15 sessions. All 15 sessions (yellow bars) were admitted for each of the bandwidth scenarios in the \emph{RA Only} architecture as shown in Figures \ref{session_mos_2M}-\ref{session_mos_9M}. Wheres, a new session will only be admitted in the \emph{Cross-Layer} architecture if there is enough bandwidth. This procedure ensures that the new session does not penalize the available sessions and it will be given the required QoE. Figure \ref{session_bandwidth} explains the number of the admitted sessions of both architectures for each bandwidth configuration. The number of the admitted sessions increased with the increase of the bandwidth in the \emph{Cross-Layer} architecture although still remains less than the \emph{RA Only} architecture.

The admission decision of the admission procedure in this paper was based on the instantaneous aggregate arrival rate. In another paper \cite{Qadir2013a}, we have shown that the average aggregate arrival rate is a better efficient decision parameter than the instantaneous aggregate arrival rate to be taken for few number of video sessions. This means that using the average aggregate arrival rate guarantees more sessions to be admitted for as few number as 15 sessions. 

Figures \ref{session_mos_2M}-\ref{session_mos_9M} show the MOS of the admitted sessions of both architectures for different bandwidth scenarios. All admitted sessions were guaranteed at least the score of 4 (good) by the \emph{Cross-Layer} architecture. While the majority of the 15 admitted sessions were scored (bad) or (poor) in the \emph{RA Only} architecture due to the dispersion of quality among the admitted sessions. Furthermore, higher bandwidth of the bottleneck link in the \emph{Cross-Layer} architecture guarantees the QoE of the new admitted session while keeping the QoE of the available sessions.

Table \ref{table:drop_utilization} summarizes the percentage of the byte drop ratio and link utilization for both architectures. The overall trend of the drop is lower and utilization is higher for the \emph{Cross-Layer} architecture due to better bandwidth management by the admission control procedure and more efficient utilization of the link. More sessions which compete for the available bandwidth will be accommodated whenever more bandwidth becomes available. It therefore, causes more drop and less utilization; however, the drop is still less than 1\% and utilization is considerably higher. In contrast to the \emph{Cross-Layer} architecture, the decrease of the drop caused by higher bandwidth in the \emph{RA Only} architecture doesn't make a better use of the link as seen from the utilization figures of the table. Hence, increasing the bandwidth does not improve the utilization and doesn't take the QoE of all sessions to an acceptable level (the MOS of 3 which is considered a fair quality) due to the fact that more sessions will be admitted which share the available bandwidth and discriminate the active sessions.

\section{Conclusion}
A QoE-aware cross-layer architecture for video traffic over the Internet was proposed in this paper to deal with the extreme growth of the video traffic on Internet in the foreseeable future. A combination of the rate-adaptation and admission control functionalities were modeled in the proposed architecture. The proposed cross-layer architecture was compared to an architecture in which video sources adapting their transmission rate only using simulation. The simulation results have shown that the cross-layer architecture guarantees the QoE of the admitted sessions and utilizes the link more efficiently comparing to the rate adaptation only architecture. Future work will introduce an admission control based on the average aggregate arrival rate. Other QoE indexes such as the Structural SIMilarity (SSIM) will be used to evaluate the quality of the decoded video.

\begin{table}[!th]
\centering
\caption{The drop ratio and utilization of cross-layer and RA only architectures}
\label{table:drop_utilization}
\begin{tabular}{|c|c|c|c|c|}
                                                                         \hline
   & \multicolumn{2}{|c|}{{\emph{Cross-Layer}}}  &    	\multicolumn{2}{|c|}{{\emph{RA Only}}} \\\hline   
Bandwidth(Mbps) & Drop \% & Utilization & Drop \% & Utilization\\\hline                                           
2 & 0.68 & 91.02 & 3.66 & 87.11 \\\hline
4 & 0.55 & 88.84 & 1.55 & 80.01 \\\hline
6 & 0.67 & 86.51 & 1.29 & 72.50 \\\hline
9 & 0.82 & 83.77 & 0.57 & 62.39 \\\hline
\end{tabular}
\end{table}



\end{document}